\documentstyle[12pt]{article}    
\renewcommand{\theequation}{\thesection.\arabic{equation}}
\begin{document}
\setlength{\baselineskip}{0.2in}
\begin{titlepage}
\noindent 

\begin{flushright}
\begin{tabular}{l}
nucl-th/9609001\\
Revised July. 1997\\
\end{tabular}
\end{flushright}

\noindent
\vspace{36pt}

\vspace{5pt}

\centerline{\Large{\bf QCD sum rules for $\rho$, $\omega$, $\phi$
meson-nucleon scattering}}
\centerline{\Large{\bf lengths and the mass shifts in nuclear medium}}

\par             
\par\bigskip
\par\bigskip
\vspace{0.7cm}                                                 

\centerline{Yuji Koike and A. Hayashigaki}
\centerline{\it  Graduate School of Science and Technology,
Niigata University, Ikarashi, Niigata 950-21, Japan}
\vspace{1in}

\centerline{\bf Abstract}
A new QCD sum rule analysis on
the spin-isospin averaged
$\rho$, $\omega$ and $\phi$ meson-nucleon scattering lengths 
is presented.
By introducing the constraint relation on the low 
energy limit of the
vector-current nucleon forward scattering amplitude
(low energy theorem), we 
get $a_\rho=-0.47\pm 0.05$ fm, $a_\omega=-0.41\pm 0.05$ fm and
$a_\phi=-0.15\pm 0.02$ fm, which suggests that these 
$V-N$ interactions
are attractive.
It is also proved that the previous studies on the 
mass shift of these vector mesons in the nuclear medium 
are essentially 
the ones obtained from these scattering lengths
in the linear density approximation.

\vspace{1.0cm}


\end{titlepage}

\newpage
\section{Introduction}
\setcounter{equation}{0}
\renewcommand{\theequation}{\arabic{section}.\arabic{equation}}

Modifications of hadron properties in 
nuclear medium is of great interest 
in connection with the ongoing experimental plans
at CEBAF and RHIC etc.
Especially, the mass shift of vector mesons
is directly accessible by inspecting the change 
of the lepton pair spectra
in the electro- or photo- production 
experiments of the vector mesons from the 
nuclear targets.  
To study this issue,
Hatsuda-Lee (HL) applied the
QCD sum rule (QSR) method to the
vector mesons 
in the nuclear medium, and got 10-20 \% {\it decrease} 
of the masses of the $\rho$ and $\omega$ mesons at 
the nuclear matter
density\,\cite{HL}.  
Later one of the present authors\,\cite{Koike}
reexamined the analysis of \cite{HL} 
based on the observation that their density effect
in the vector current
correlator
comes from 
the current-nucleon forward
scattering amplitude, 
and accordingly the effect should be interpretable
in terms of the 
physical effect in the forward amplitude\,\cite{KM}.
This analysis showed
slight {\it increase} of the $\rho$, $\omega$ meson masses
in contradiction to \cite{HL}.
Subsequently,
the analysis in \cite{Koike} was criticized 
by Hatsuda-Lee-Shiomi\cite{HLS}.
This paper is prepared as a reexamination 
and a more expanded discussion of \cite{Koike}.  
We present a new analysis on the $\rho$, $\omega$ and $\phi$-
nucleon scattering lengths.  By introducing a constraint relation
among the parameters in the spectral function, we eventually got
a decreasing mass similar 
to \cite{HL}, although the interpretation presented in \cite{Koike}
essentially persists.
We also provide informative comments and replies to \cite{HLS},
and clarify the misunderstanding in the
literature on the interpretation of the mass shift
\,\cite{HL,HLS,Hatsuda}.

We first wish to give a brief sketch
of the debate.
The information about the spectrum of a vector meson 
in the nuclear medium 
with the nucleon density $\rho_N$
can be extracted from the correlation function
\begin{eqnarray}
\Pi_{\mu\nu}^{\rm NM}(q) = i\int d^{4}x 
e^{iq \cdot x}\langle TJ_{\mu}(x)
J_{\nu}^{\dag}(0) \rangle _{\rho_N},
\label{eq1.1}
\end{eqnarray}
where $q=(\omega,\mbox{\boldmath $q$})$ is the four momentum and 
$J_\mu$ denotes the vector current for the vector
mesons in our interest:
\begin{eqnarray}
J_{\mu}^{\rho}(x) = \frac{1}{2}
(\overline{u}\gamma_{\mu}u-\overline{d}\gamma_{\mu}d)(x),\ 
J_{\mu}^{\omega}(x) = \frac{1}{2}
(\overline{u}\gamma_{\mu}u+\overline{d}\gamma_{\mu}d)(x),\ 
J_{\mu}^{\phi}(x) = \overline{s}\gamma_{\mu}s(x).
\label{eq1.2}
\end{eqnarray}
Following a common wisdom in the QSR method\,\cite{SVZ}, 
Hatsuda-Lee
applied an operator product expansion (OPE) 
to this correlator at large $Q^2 = -q^2 > 0$.
The basic assumption employed in this procedure 
is that the $\rho_N$-dependence of the correlator 
is wholely ascribed to the $\rho_N$ dependence in the
condensates\,\cite{DL}: 
\begin{eqnarray}
\Pi^{\rm NM}(q^{2}\rightarrow -\infty) \stackrel{\rm OPE}{=}
 \sum_{i}C_{i}(q^{2},\mu^{2})
\langle {\cal O}_{i}(\mu^{2})\rangle _{\rho_N},
\label{eq1.3}
\end{eqnarray}
where $C_i$ is the Wilson coefficient
for the operator ${\cal O}_i$ and we suppressed 
all the Lorentz indices
for simplicity.
A new feature in the finite density sum rule 
is that both Lorentz scalar and nonscalar operators
survive as the condensates $\langle {\cal O}_i \rangle_{\rho_N}$.
An assumption of the Fermi gas model for the nuclear medium
was introduced to estimate the $\rho_N$-dependence of
$\langle {\cal O}_i \rangle_{\rho_N}$, which is 
expected to be valid at
relatively low density\,\cite{DL}:
\begin{eqnarray}
\langle {\cal O}_i\rangle_{\rho_N} 
&=& \langle {\cal O}_i \rangle_0
+ \sum_{\rm spin, isospin}\int^{p_f} 
{ d^3p \over  (2\pi)^3 2p^0 }
\langle ps|{\cal O}_i | ps \rangle \nonumber\\
&=& \langle {\cal O}_i\rangle _{0} + 
\frac{\rho_N}{2M_{N}}\langle {\cal O}_i\rangle _{N} + o(\rho_N),
\label{eq1.4}
\end{eqnarray}
where $\langle\cdot\rangle_0$ represents the 
vacuum expectation value,
$|ps\rangle$ denotes the nucleon state with momentum $p$
and the spin $s$
normalized covariantly 
as $\langle ps|p's'\rangle = (2\pi)^3 2p^0 \delta_{ss'}
\delta^{(3)}(\mbox{\boldmath $p$}
-\mbox{\boldmath $p$}')$, and $\langle \cdot\rangle_N$
denotes the expectation value with respect to the nucleon state
with $\mbox{\boldmath $p$}=0$.
The effect of $\mbox{\boldmath $p$} 
\neq 0$ introduces $O(\rho_N^{5/3})$ correction to
(\ref{eq1.4}).
This way the $\rho_N$-dependence of the condensates 
can be incorporated through the nucleon matrix elements
in the linear density approximation.
By inserting (\ref{eq1.4}) in (\ref{eq1.3}), one can easily see that
the approximation to the condensate,
(\ref{eq1.4}), is equivalent to the following approximation to
the correlation function itself:
\begin{eqnarray}
\Pi^{\rm NM}_{\mu\nu}(q) = \Pi^0_{\mu\nu}(q)
+ \sum_{\rm spin, isospin}\int^{p_f} { d^3p \over (2\pi)^3 2p^0 }
T_{\mu\nu}(p,q),
\label{eq1.5}
\end{eqnarray}
where $\Pi^0_{\mu\nu}(q)$ is the vector current correlator
in the vacuum,
\begin{eqnarray}
\Pi^0_{\mu\nu}(q)=i\int d^{4}x e^{iq \cdot x}\langle  
\mbox{T}J_{\mu}(x)
J_{\nu}^{\dag}(0) \rangle_0,
\label{eq1.6}
\end{eqnarray}
and
$T_{\mu\nu}(p,q)$ is the current-nucleon forward amplitude
defined as
\begin{eqnarray}
T_{\mu\nu}(p,q)= i \int d^4x e^{iq\cdot x}\langle ps | 
T J_\mu(x)J^{\dag}_\nu (0)|ps \rangle.
\label{eq1.7}
\end{eqnarray}
Since \cite{HL} adopted (\ref{eq1.4}), 
one should be able to interpret the result in \cite{HL}
from the point of view of the current-nucleon forward amplitude.
What was the essential ingredient 
in $T_{\mu\nu}$ which lead to the decreasing 
mass in \cite{HL}?   What kind of approximation 
in the analysis of $T_{\mu\nu}(p,q)$ corresponds to
the analysis of $\Pi_{\mu\nu}^{\rm NM}$ in \cite{HL}?    

To answer these questions 
we first note that the linear density approximation
(\ref{eq1.4}) to the condensates becomes better at smaller $\rho_N$
or equivalently smaller $p_f$.  As long as the OPE side 
is concerned,
the effect of the nucleon Fermi motion can be
included in $\langle{\it O}\rangle_{\rho_N}$
as is discussed in \cite{HLS}.  
It turned out, however, that its effect is 
negligible.
Therefore what is relevant in the mass shift in the QSR approach 
is the structure of $T_{\mu\nu}$ in the 
$\mbox{\boldmath $p$}=0$ limit. 
We observe that in this limit,
$T_{\mu\nu}$
is reduced to the vector meson-nucleon scattering length $a_V$
at $q=(\omega=m_V,\mbox{\boldmath $q$}=0)$ 
($m_V$ is the mass of the vector meson).
If one knows
$a_V$, the mass shift of the vector meson becomes
\begin{eqnarray}
\delta m_{V} = 2\pi \frac{M_{N}+m_{V}}{M_{N}m_{V}}a_{V }\rho_N
\label{eq1.8}
\end{eqnarray}
in the linear density approximation.  In the following discussion
we argue that what
was observed in \cite{HL} as a decreasing mass shift is essentially
the one in (\ref{eq1.8}).
Of course, whether the approximation (\ref{eq1.4}), (\ref{eq1.5})
to $\Pi^{\rm NM}_{\mu\nu}$
is a good one or not  
at the nuclear matter density is a different issue.
What we wish to stress is that the approximation
adopted in \cite{HL} is certainly interpretable in terms of the 
vector meson-nucleon ($V-N$)
scattering lengths unlike the argument in \cite{HLS}.

To motivate our idea from a purely mathematical point of view,
let's forget about the $V-N$ scattering lengths
for the moment, and translate what was observed in \cite{HL}
into the language of $T_{\mu\nu}$. 
HL analyzed $\Pi^{\rm NM}_1(\omega^2)\equiv 
\Pi_\mu^{{\rm NM}\mu}(q)/(-3\omega^2)$
at $\mbox{\boldmath $q$}=0$ in QSR. At $\rho_N=0$, 
namely in the vacuum,
$\Pi^{\rm NM}_1(\omega^2)$ is reduced to $\Pi_1(q^2)$ 
defined by the relation
$\Pi^0_{\mu\nu}(q)=(q_\mu q_\nu -g_{\mu\nu}q^2)\Pi_1(q^2)$.
HL obtained a QSR relation for $\Pi_1^{\rm NM}$ as
\begin{eqnarray}
{1 \over 8\pi^2}{\rm ln}\left( {s_0^* - q^2 \over -q^2 } \right)
+{ A^* \over q^4} + { B^* \over q^6} = { F'^* \over m_V^{2*}-q^2}
+ { \rho_{sc} \over q^2},
\label{eq1.9}
\end{eqnarray}
where $A^*$ and $B^*$ are 
the in-medium condensates with dim.=4 and dim.=6, respectively,
and $m_V^{*2}$, $F^*$ and $s_0^*$ are the in-medium values
of the (squared) vector meson mass, 
pole residue and the continuum threshold,
which are to be determined by fitting the above equation.
$\rho_{sc}$ is the so called Landau damping term
which is purely a medium effect and is thus 
$O(\rho_N)$. Actual values are 
$\rho_{sc}=-{\rho_N \over 4M_N}$ for the $\rho$, 
$\omega$ mesons and $\rho_{sc}=0$ for $\phi$ meson\,\cite{HL,BS}.
At $\rho_N=0$, (\ref{eq1.9}) is simply the well known sum rule in the
vacuum\,\cite{SVZ}:
\begin{eqnarray}
{1 \over 8\pi^2}{\rm ln}\left( {s_0 - q^2 \over -q^2 } \right)
+{ A \over q^4} + { B \over q^6} = { F' \over m_V^2-q^2}.
\label{eq1.10}
\end{eqnarray}
Since HL
included the linear density correction (\ref{eq1.4}) 
in $A^*$ and $B^*$,
they got the change in $m_V^{*2}$, $F^*$ and $s_0^*$
to $O(\rho_N)$ accuracy.
Indeed, HL got a clear linear change in these quantities.
We write $A^*=A+{\rho_N \over 2M_N}\delta A$ and
similarly for $B^*$ corresponding to (\ref{eq1.4}), where
$\delta A$ and $\delta B$ are the nucleon matrix elements
of the same operators as $A$ and $B$ respectively.
Correspondingly it is legitimate to write
$m_V^{2*}=m_V^2+{\rho_N \over 2M_N}\delta m_V^2 $, 
$F'^*=F'+{\rho_N \over 2M_N}\delta F'$ and 
$s_0^*=s_0+{\rho_N \over 2M_N}\delta s_0$.
Expand (\ref{eq1.9}) to $O(\rho_N)$ and subtract
(\ref{eq1.10}) from it.  Then one gets
\begin{eqnarray}
{ \delta A \over q^4} + {\delta B \over q^6} =
{-F'\delta m_V^{2} \over (m_V^2 - q^2)^2} +{\delta F' 
\over m_V^2 -q^2}
-{\delta s_0/(8\pi^2) \over s_0 -q^2 } + 
{\delta\rho_{sc} \over q^2}. 
\label{eq1.11}
\end{eqnarray}
The left hand side of this equation is precisely the 
OPE expression for 
$T_\mu^\mu (p,q)/(-3\omega^2)$ 
at $\mbox{\boldmath $p$}=\mbox{\boldmath $q$}=0$, and thus 
(\ref{eq1.11}) is the QSR for the same quantity which 
is equivalent to 
the QSR for $\Pi_1^{\rm NM}(\omega^2)$ assumed in \cite{HL}.
Regardless of what HL intended in their
sum rule analysis for the the vector mesons in the medium,
(\ref{eq1.11}) is the equivalent sum rule relation for
$T_{\mu\nu}$ in their analysis.
What is the physical content of this 
sum rule for $T_{\mu\nu}$?
In this paper 
we shall show that our analysis on the vector meson
nucleon scattering 
lengths precisely leads to the sum rule (\ref{eq1.11}).

This paper is organized as follows.
In section 2, we present 
a new analysis for the $\rho$, $\omega$ and $\phi$ 
meson-nucleon spin-isospin averaged scattering lengths
in the framework of QSR.  The difference from 
the previous analysis\,\cite{Koike} is emphasized.
The contents of this section
should be taken as independent from the issue of the mass shift
of these vector mesons in the nuclear medium.
In section 3, we discuss the relation between the scattering lengths
obtained in section 2 and the mass shift of \cite{HL}.  In section 4,
we shall
give detailed answers and comments to the criticisms raised in 
\cite{HLS}.
Section 5 is devoted to summary and conclusion.  Some of the formula
will be discussed in the appendix.

\section{$\rho$, $\omega$, $\phi$-nucleon scattering lengths}
\renewcommand{\theequation}{\arabic{section}.\arabic{equation}}
\setcounter{equation}{0}

In this section we analyze 
the vector current-nucleon forward scattering
amplitude (\ref{eq1.7}) 
at $\mbox{\boldmath $p$}=0$ in the framework of the QCD sum rule,
and present a new estimate for the $\rho$, $\omega$ and $\phi$-meson
nucleon scattering lengths.  
We first write
\begin{eqnarray}
T_{\mu\nu}(\omega,\mbox{\boldmath $q$}) 
& = & i\int d^{4}x e^{iq \cdot x}
\langle ps| \mbox{T}J_{\mu}(x)J_{\nu}^{\dag}(0) |ps\rangle,
\label{eq2.1}
\end{eqnarray}
suppressing the explicit dependence on
the four momentum of the nucleon $p=(M_N,0)$.
As was noticed in the introduction, 
we are interested in the structure
of $T_{\mu\nu}(\omega,\mbox{\boldmath $q$}=0)$ 
around $\omega=m_V$ which 
affects the pole structure of the vector current 
correlator in the medium.
Near the pole position of the vector meson, 
$T_{\mu\nu}$ can be associated 
with the $T$ matrix for the forward $V-N$ ($V=\rho,\omega, \phi$)
scattering amplitude 
${\cal T}_{hH,h'H'}$
by the following relation
\begin{eqnarray}
\epsilon^{*\mu}_{(h)}(q)T_{\mu\nu}(\omega,\mbox{\boldmath $q$})
\epsilon^{\nu}_{(h')}(q) \simeq
\frac{-f_{V}^{2}m_{V}^{4}}{(q^{2}-m_{V}^{2}+i\varepsilon)^{2}} 
{\cal T}_{hH,h'H'}(\omega,\mbox{\boldmath $q$}),
\label{eq2.2}
\end{eqnarray}
where $h$($h'$) denotes the helicities for the initial (final) 
vector meson, and similarly 
$H$($H'$) for the nucleon.   In (\ref{eq2.2}) the coupling $f_V$
is introduced  
by the relation
$\langle 0|J_{\mu}^{V}|V^{(h)}(q)\rangle =  
f_{V}m_{V}^{2}\epsilon_{\mu}^{(h)}(q)$
with the 
polarization vector $\epsilon_\mu^{(h)}$ normalized as
$\sum_{h}\epsilon_{\mu}^{(h)^{*}}(q)\epsilon_{\nu}^{(h)}(q)
=-g_{\mu\nu}+q_{\mu}q_{\nu}/q^{2}$.
$T_{\mu\nu}$ can be decomposed into the four scalar functions
respecting the invariance under parity and time reversal and the
current conservation.
Taking the spin average on both sides of
(\ref{eq2.2}) (see appendix A), 
$T_{\mu\nu}(\omega,\mbox{\boldmath $q$})$ is projected onto
$T(\omega,\mbox{\boldmath $q$})=
T_\mu^\mu/(-3)$\,
\cite{err} and 
${\cal T}_{hH,h'H'}$ is projected onto
the spin averaged $V-N$ $T$ matrix, ${\cal T}(\omega, 
\mbox{\boldmath $q$})$.
At $q=(m_V,0)$ and $p=(M_N,0)$, ${\cal T}$ is connected to
the spin averaged $V-N$ scattering length $a_V$  
as ${\cal T}(m_V, 0)=
8\pi (M_N+m_V)a_V$\,\cite{err} 
with $a_V={1 \over 3}(2a_{3/2}+a_{1/2})$ 
where $a_{3/2}$ and $a_{1/2}$
are the $V-N$ scattering lengths in the spin-3/2 and 1/2
channels respectively.  We also remind that the $\rho^0-N$
 scattering
length corresponds to the 
isospin-averaged scattering length owing to the
isospin symmetry.

The retarded correlation function defined by 
\begin{eqnarray}
T_{\mu\nu}^{R}(\omega,\mbox{\boldmath $q$})
 = i\int d^{4}x e^{iq \cdot x}
\langle N|\theta(x^{0}) [J_{\mu}(x),J_{\nu}^{\dag}(0)]
 |N\rangle 
\label{eq2.2p}
\end{eqnarray}
satisfies the following dispersion relation
\begin{eqnarray}
T_{\mu\nu}^{R}(\omega,\mbox{\boldmath $q$}) 
= \frac{1}{\pi}\int_{-\infty}^{\infty}du
\frac{\mbox{Im}\ T^{R}_{\mu\nu}
(u,\mbox{\boldmath $q$})}{u-\omega-i\varepsilon}.
\label{eq2.3}
\end{eqnarray}
We recall that
for nonreal values of $\omega$, 
$T^R_{\mu\nu}(\omega, \mbox{\boldmath $q$})$ 
becomes identical to 
$T_{\mu\nu}(\omega,\mbox{\boldmath $q$})$.
Applying the same spin-averaging 
procedure to both sides of (\ref{eq2.3}) as above,
we get the following dispersion relation for $\omega^2\neq$
positive real number:
\begin{eqnarray}
T(\omega,0)= \int_{-\infty}^\infty\, d\,u{ \rho(u, 0)
\over u-\omega -i\epsilon} = \int_0^\infty\,d\,u^2 {\rho(u,0)
\over u^2 -\omega^2 },
\label{eq2.4}
\end{eqnarray}
where we introduced the 
spin-averaged spectral 
function $\rho(\omega,\mbox{\boldmath $q$})$
constructed from ${1\over \pi}{\rm Im}
T_{\mu\nu}^R(\omega,\mbox{\boldmath $q$})$.
The second equality in (\ref{eq2.4}) comes from the relation
$\rho(-\omega,-\mbox{\boldmath $q$})=-\rho(\omega,
\mbox{\boldmath $q$})$.
Using (\ref{eq2.2}),
$\rho(u, 0)$
can be expressed 
in terms of the spin-averaged $V-N$ forward $T$-matrix 
${\cal T}$ as
\begin{eqnarray}
\lefteqn{\rho(u>0,\mbox{\boldmath $q$}=0)} \nonumber \\
 &=& \frac{1}{\pi}\mbox{Im} \left[
\frac{-f_{V}^{2}m_{V}^{4}}{(u^{2}-m_{V}^{2}
+i\varepsilon)^{2}}{\cal T}(u,0) 
\right]+\cdots\nonumber \\
&=& \frac{-f_{V}^{2}m_{V}^{4}}{\pi}\left[
\mbox{Im} \frac{1}{(u^{2}-m_{V}^{2}+i\varepsilon)^{2}} \mbox{Re} 
{\cal T}(u,0) 
+ \mbox{Re} \frac{1}{(u^{2}-m_{V}^{2}+i\varepsilon)^{2}} 
\mbox{Im} {\cal T}(u,0)
\right] +\cdots\label{eq2.5} \\
&\equiv & a\delta'(u^{2}-m_{V}^{2}) + b\delta(u^{2}-m_{V}^{2})
 + c\delta(u^{2}-s_{0}),
\label{eq2.6}
\end{eqnarray}
where
\begin{eqnarray}
a &=& -f_V^2 m_V^4 {\rm Re}{\cal T}(u,0)|_{u=m_V} = -8\pi f_V^2
m_V^4(M_N +m_V)a_V,\label{eq2.6p}\\
b &=& -f_V^2 m_V^4 { d \over du^2}{\rm Re}{\cal T}(u,0)|_{u=m_V},
\label{eq2.6pp}
\end{eqnarray}
and $\cdots$ in (\ref{eq2.5}) represents the 
continuum contribution which is not associated with
the $\rho-N$ scattering.\footnote{In (\ref{eq2.2}),
${\cal T}$ is defined only around $\omega=m_V$ and thus we
introduced the contribution $\cdots$ in (\ref{eq2.5}).}
The first two terms in (\ref{eq2.6}) come from the first term in
(\ref{eq2.5}) when (\ref{eq2.5}) is substituted into the dispersion
integral (\ref{eq2.4}).
The $b$-term (simple pole term) in (\ref{eq2.6}) represents
the off-shell effect in the ${\cal T}$ matrix of the forward 
$VN\to VN$ scattering.
We note that no other higher
derivatives of ${\rm Re}{\cal T}(u,0)$ appear here. 
The third term in (\ref{eq2.6})
corresponds to $\cdots$ in (\ref{eq2.5}) and represents the 
scattering contribution in the continuum part of $J_V$ which starts at 
the threshold $s_0$.  The value of $s_0$ is fixed as $s_0=1.75$ GeV$^2$
for the $\rho$ and $\omega$ mesons 
and $s_0= 2.0$ GeV$^2$ for the $\phi$ meson,
since these values are known to reproduce the masses of these 
mesons\,\cite{SVZ}.
What is not included in the ansatz (\ref{eq2.6}) 
is the second term in $[\cdots ]$ of (\ref{eq2.5}) which represents 
inelastic (continuum) 
contribution such as $\rho N\to \pi N, \pi \Delta$ for
the $\rho$ meson and $\phi N\to K\Lambda, 
K\Sigma$ for the $\phi$ meson etc.
The strength of
these contributions could be sizable, so we should take the following
analysis with caution. (See discussion below.)

The OPE expression for $T(q^2=\omega^2)=T(\omega,0)$ in (\ref{eq2.4})
is given in Eq. (6) of \cite{Koike} 
(and Eq. (2.13) of \cite{HKL}) for the $\rho$ and $\omega$-mesons,
and it is not repeated here.  It takes the following form
including the operators with dimension up to 6:
\begin{eqnarray}
T^{\rm OPE}(q^2) = { \alpha \over q^2} + {\beta \over q^4},
\label{eq2.7}
\end{eqnarray}
where $\alpha$ is 
the sum of the nucleon matrix elements of the dim.=4 operators
and $\beta$ for the 
dim.=6 operators.  In our analysis, we adopt the same values 
for these matrix elements as \cite{Koike}:  $\alpha= 0.39$ GeV$^2$
for the $\rho$ and $\omega$ 
mesons and $\beta = -0.23 \pm 0.07$ ($-0.16 \pm
0.10$) GeV$^4$ 
for the $\rho$ ($\omega$) mesons.  The difference in $\beta$
between $\rho$ and $\omega$
originates from the twist-4 matrix elements for which we adopted the 
parameterization used in \cite{CHKL}.
For the $\phi$ meson,
$\alpha=0.24$ GeV$^2$ and $\beta=-0.12$ GeV$^4$.  See \cite{Koike}
for the detail.

Up to now our procedure 
for analyzing $T_{\mu\nu}$ is completely the same
as \cite{Koike}.  Here we start to deviate from \cite{Koike}
and introduce a constraint relation
among $a$, $b$ and $c$ which is imposed by the low energy theorem
for the vector current-nucleon forward scattering amplitude.
In the low energy limit, 
$p\rightarrow (M_N, 0)$ and $q=(\omega, \mbox{\boldmath $q$})
\rightarrow (0,0)$, $T_{\mu\nu}(\omega, \mbox{\boldmath $q$})$ 
is determined
by the Born diagram contribution (Fig.1)
as in the case of the Compton scattering\,\cite{Bj}.  
Since we are considering the 
case $\mbox{\boldmath $q$}=0$, 
we first put $\mbox{\boldmath $q$}=0$
and then take the limit $\omega\to 0$ (See appendix B):
\begin{equation}
T^{\rm Born}(\omega^2)\equiv T^{\rm Born}(\omega,0) = \left\{
\begin{array}{ll} \frac{-2M_{N}^{2}}{4M_{N}^2-\omega^2} 
\stackrel{\omega\rightarrow 0}{\longrightarrow} -\frac{1}{2}
 & (\rho^{0},\omega) \\
\hspace*{3mm}0 & (\phi)
\end{array}
\right.
\label{eq2.9}
\end{equation}
At $q_{\mu} \neq 0$, the Born term is not the total contribution
and there remains an ambiguity
in dealing with $T^{\rm Born}$.
We thus assume two forms of the parameterization
for the phenomenological side of 
the sum rules for $\rho$ and $\omega$ mesons:
\begin{enumerate}
\item[(i)] With explicit Born term:
\begin{eqnarray}
T^{\rm  ph}(q^{2}) = T^{\rm Born}(q^{2}) + \frac{a}
{(m_{V}^{2}-q^2)^{2}}
+ \frac{b}{m_{V}^{2}-q^2}
+ \frac{c}{s_{0}-q^2} 
\label{eq2.10}
\end{eqnarray}
with the condition
\begin{eqnarray}
\frac{a}{m_{V}^{4}}
+ \frac{b}{m_{V}^{2}}
+ \frac{c}{s_{0}} = 0.
\label{eq2.11}
\end{eqnarray}
\item[(ii)] Without explicit Born term:
\begin{eqnarray}
T^{\rm ph}(q^{2}) = \frac{a}{(m_{V}^{2}-q^2)^{2}}
+ \frac{b}{m_{V}^{2}-q^2}
+ \frac{c}{s_{0}-q^2}
\label{eq2.12}
\end{eqnarray}
with the condition 
\begin{eqnarray}
\frac{a}{m_{V}^{4}}
+ \frac{b}{m_{V}^{2}}
+ \frac{c}{s_{0}} = T^{\rm Born}(0).
\label{eq2.13}
\end{eqnarray}
\end{enumerate}
With the phenomenological sides of the sum rules 
((\ref{eq2.10}) or (\ref{eq2.12})) and the OPE side (\ref{eq2.7}), 
the QSR is given by the relation
\begin{eqnarray}
T^{\rm OPE}(q^2) = T^{\rm ph}(q^2).
\label{QSR}
\end{eqnarray}
Several comments are in order here.
\begin{enumerate}

\item
Because of the conditions 
(\ref{eq2.11}) and (\ref{eq2.13}), $T^{\rm ph}(q^2)$
satisfies $T^{\rm ph}(0)=T^{\rm Born}(0)$ and 
has two independent parameters to be determined in either case.
This part is the essential difference from the previous study
in \cite{Koike}.  In \cite{Koike}, $a$, $b$ and $c$ were treated as 
independent parameters which were determined 
in the Borel sum rule (BSR).
In the following, we eliminate $c$ by these relations and regard
$T^{\rm ph}$ as a functions of $a$ and $b$.

\item 
The leading behavior of $T^{\rm ph}(q^2)$
at large $-q^2 >0$ is consistent with $T^{\rm OPE}(q^2)$:
Both sides start with the ${1\over q^2}$ term, which supports
the form of the spectral function in (\ref{eq2.6}).

\item
Inclusion of $T^{\rm Born}(q^2)$ in (\ref{eq2.10}) has a 
similar effect
as the inclusion of the ``second continuum'' contribution
with the threshold $4M_N^2$.  In the QSR analysis for 
the lowest resonance
contribution, the result is more reliable if it does not depend
on the details of the higher energy part.
We shall see this is indeed the case in the 
following Borel sum rule method.

\end{enumerate}

By expanding $T^{\rm ph}(q^2)$ with respect to
$1/(-q^2)$ and comparing the coefficients 
of $1/q^2$ and $1/q^4$ in $T^{\rm ph}(q^2)$ 
with those in $T^{\rm OPE}(q^2)$, one gets the
finite energy sum rules (FESR).  These relations are solved to give
\begin{eqnarray}
a &=& { 1 \over 1- { s_0 \over m_V^2}} \left[ m_V^2 \left( 
1 + {s_0 \over m_V^2}\right)\left(-\alpha + 2M_N^2\right)
+ \left( \beta - 8 M_N^4 \right) \right],
\label{eq2.14}\\
b &=& { 1 \over \left( 1 - {s_0 \over m_V^2} \right)^2 }
\left[ \left(1 + { s_0^2 \over m_V^4 } \right) 
\left( -\alpha + 2 M_N^2 
\right)
 + { s_0 \over m_V^4 } \left( \beta - 8M_N^4 \right) \right], 
\label{eq2.15}
\end{eqnarray}
for the case (i) and
\begin{eqnarray}
a &=& { 1 \over 1- { s_0 \over m_V^2}} \left[ m_V^2 \left( 
1 + {s_0 \over m_V^2}\right)\left(-\alpha + { 1 \over 2}s_0 \right)
+ \left( \beta - {1 \over 2}s_0^2 \right) \right],
\label{eq2.16}\\
b &=& { 1 \over \left( 1 - {s_0 \over m_V^2} \right)^2 }
\left[ \left(1 + { s_0^2 \over m_V^4 } \right) 
\left(-\alpha + { 1 \over 2}s_0
\right)
 + { s_0 \over m_V^4 } \left( \beta - 
{ 1 \over 2}s_0^2 \right) \right],
\label{eq2.17}
\end{eqnarray}
for the case (ii).
These FESR relations give $a_{\rho}=-0.68$ fm, 
$a_{\omega}=-0.66$ fm
 for the case (i) and $a_{\rho}=-0.13$ fm, $a_{\omega}=-0.11$ fm
for the case (ii).
For the $\phi$ meson, $a_{\phi}=-0.06$ fm.  
Two ways of dealing with the Born term
give quite different results.  This is not surprising.
Since the leading order contribution in $T^{\rm ph}$ comes from 
the continuum contribution, the results in FESR strongly depends on 
the treatment of this part.  
These small negative numbers, 
however, suggest that the $V-N$ interaction
is weakly attractive.

In order to give more 
quantitative prediction, we proceed to the Borel
sum rule (BSR) analysis.  
In this method, the higher energy contribution
in the spectral function is suppressed compared 
to the $V-N$ scattering
contribution.  We thus have an advantage that the ambiguity in
dealing with the Born term becomes less important in BSR.  
We shall try the following two methods in BSR:

\begin{enumerate}
\item[(1)] Derivative Borel Sum Rule (DBSR):
After the Borel transform of (\ref{QSR})
with respect to $Q^2=-q^2>0$, take the derivative of both sides
with respect to the Borel mass $M^2$, and use those two equations
to get $a$ and $b$ by taking the average in a Borel window,
$M_{min}^2<M^2<M_{max}^2$.

\item[(2)] Fitting Borel Sum Rule (FBSR):
Determine $a$ and $b$ in order to make the 
following quantity minimum
in a Borel window $M_{\rm min}^2 < M^2 < M^2_{\rm max}$:
\begin{equation}
F(a,b)=\int_{M_{min}^{2}}^{M_{max}^{2}}dM^{2}[T^{\rm OPE}(M^{2})
- T^{\rm ph}(M^{2};a,b)]^{2}
\label{eq2.18}
\end{equation}
where $T^{\rm ph}(M^2; a, b)$ is the Borel transform 
of $T^{\rm ph}(q^2)$ which is a functional of $a$ and $b$.

\end{enumerate}

After getting $a$ and $b$ by these methods, we determine
$a_V$ from the relation (\ref{eq2.6p}) using the 
experimental values 
of $M_N$, $m_V$ and $f_V$.  The numbers we adopted are
$M_N=940$ MeV, $m_{\rho,\omega}=770$ MeV, $f_{\rho,\omega}=0.18$,
$m_{\phi}=1020$ MeV and $f_{\phi}=0.25$.
Borel curves for $a_V$ ($V$=$\rho$, $\omega$, $\phi$) in the DBSR
are shown in Figs. 2 and 3.   
Stability of these curves is reasonably good
around $M^2=1$ GeV$^2$ for the $\rho$ and $\omega$ mesons
and around $M^2=1.5$ GeV$^2$ for the $\phi$ meson.
We take the average over the window $0.8\ {\rm GeV}^2\ < M^2 <
1.3\ {\rm GeV}^2$ for $\rho$, $\omega$ mesons and
$1.3\ {\rm GeV}^2\ < M^2 < 1.8\ {\rm GeV}^2$ for the $\phi$ meson.  
These windows are typical for the analysis of these vector meson
masses, and the experimental values are well reproduced
with the continuum threshold $s_0=1.75$ GeV$^2$ 
for $\rho$, $\omega$
and $s_0=2.0$ GeV$^2$ for $\phi$\,\cite{SVZ}.
The obtained values are
$a_{\rho}=-0.5$ ($-0.4$) fm 
and $a_{\omega}=-0.45$ ($-0.35$) fm for the case (i) ((ii)),
and 
$a_{\phi}=-0.15$ fm.
In the FBSR method with the same window, we get close numbers
$a_{\rho}=-0.52$ ($-0.42$) fm 
and $a_{\omega}=-0.46$ ($-0.36$) fm for (i) ((ii)),
 and $a_{\phi}=-0.15$ fm.
We tried FBSR for various Borel windows within 
$0.6$ GeV$^2$ $<M^2<1.8$ GeV$^2$
($0.9$ GeV$^2$ $<M^2<2.0$ GeV$^2$) 
for the $\rho$, $\omega$ ($\phi$) mesons and found that
the results change within 20 \% level.
From these analyses, we get
\begin{eqnarray}
a_{\rho} &=& -0.47\pm 0.05\ \mbox{fm}, \nonumber \\
a_{\omega} &=& -0.41 \pm 0.05\ \mbox{fm}, \nonumber \\
a_{\phi} &=& -0.15\pm 0.02\ \mbox{fm},
\label{eq2.19}
\end{eqnarray}
where the assigned error bars are due to the uncertainty in
the Borel analysis.

We first note that the magnitudes of these scattering lengths are
quite small, i.e., smaller than the typical hadronic size
of 1 fm.  For $\pi N$ and $K N$ systems, 
the scattering lengths 
are known to be small due to the chiral symmetry.
The above numbers 
are not so different from $a_{\pi N}$ and $a_{K N}$.
Small negative values suggest that these $V-N$ interactions
are weakly attractive.  The ansatz (\ref{eq2.6}) 
for the spectral function
ignores various inelastic contributions as was noted
below (\ref{eq2.6pp}).  So we should take the above numbers as
a rough estimate of the order of magnitude.

Recently Kondo-Morimatsu-Nishino calculated the $\pi N$ and $KN$
scattering lengths by
applying the same QSR method to the correlator of 
the axial vector current
\,\cite{KMN}.
The results with the lowest dimensional operators
in the OPE side is the same as the current algebra
calculation.  QSR supplies the correction due to the 
nucleon matrix elements of the higher
dimensional operator.  Since there is no
algebraic technique (such as current algebra)
to calculate the scattering lengths 
in the vector channels, 
it is interesting to see that OPE provides 
a possibility to estimate the strengths of the $VN$ interactions.

\section{Mass shift of the vector mesons in the nuclear medium}
\setcounter{equation}{0}
\renewcommand{\theequation}{\arabic{section}.\arabic{equation}}

In the previous section, we have identified the pole
structure of $T_{\mu\nu}(\omega,0)$ around $\omega^2 = m_V^2$ as
\begin{eqnarray}
T_{\mu\nu}(\omega,0)= 
\left( {q_\mu q_\nu \over \omega^2}-g_{\mu\nu}\right)
\left({ a \over (m_V^2 - \omega^2)^2 } 
+ { b \over m_V^2-\omega^2 } 
+ \ldots \right).
\label{eq3.0}
\end{eqnarray}
By combining this piece with the vacuum 
piece $\Pi_{\mu\nu}^0(\omega,0)$
in (\ref{eq1.6}),
the vector current correlation function in the nuclear medium
take the following form around $\omega^2=m_V^2$:
\begin{eqnarray}
\Pi_{\mu\nu}^{\rm NM}(\omega,0) &\simeq & 
\left( {q_\mu q_\nu \over \omega^2}-g_{\mu\nu}\right)
\left(\Pi(\omega^2)
+ { \rho_N \over 2M_N}T(\omega,0)\right)\nonumber\\  
&\propto & \frac{F}{m_{V}^{2}-\omega^2}+
\frac{\rho_N}{2M_{N}}\left\{\frac{a}{(m_{V}^{2}-\omega^2)^{2}} +
\frac{b}{m_{V}^{2}-\omega^2} \right\}\cdot\cdot\cdot \nonumber\\
&\simeq & \frac{F+\delta F}{(m_{V}^{2}+\Delta m_{V}^{2})-\omega^2}
 + \cdot\cdot\cdot,
\label{eq3.1}
\end{eqnarray}
where $\Pi(q^2)$ is defined as $\Pi_{\mu\nu}^0(q)=
\left( {q_\mu q_\nu \over q^2}- g_{\mu\nu}\right)\Pi(q^2)$ and
the pole residue $F$ in $\Pi^0_{\mu\nu}$
is related to $f_V$ and $m_V$ by the relation
$F=f_V^2 m_V^4$ and $\delta F={\rho_N \over 2M_N}b$.
The quantity
\begin{eqnarray}
\Delta m_{V}^{2} = \frac{-\rho_N}{2M_{N}}\frac{a}{F}
 = \frac{\rho_N}{2M_{N}}8\pi(M_{N}+m_{V})a_{V}
\label{eq3.2}
\end{eqnarray}
is regarded as the shift of the squared vector 
meson mass in nuclear matter.
We thus have the mass shift 
$\delta m_V$ as shown in (\ref{eq1.8})
from the relation
\begin{eqnarray}
m_V^*=m_{V}+\delta m_V = \sqrt{m_{V}^{2}+\Delta m_{V}^{2}}.
\label{eq3.3}
\end{eqnarray}
Using the scattering lengths obtained in the previous section,
we plotted the vector meson masses in Fig. 4 as a function
of the density $\rho_N$ based on the linear density
approximation.
At the nuclear matter density $\rho_N=0.17$ fm$^{-3}$ as
\begin{eqnarray}
\delta m_{\rho} &=& -45 \sim -55\ \mbox{MeV}\  
(6 \sim 7\%), \nonumber \\
\delta m_{\omega} &=& -40 \sim -50\ \mbox{MeV}\  
(5 \sim 6\%), \nonumber \\
\delta m_{\phi} &=& -10 \sim -20\ \mbox{MeV}\  (1 \sim 2\%).
\label{table2}
\end{eqnarray}

In order to clarify the 
relation between the above mass shifts and the
approach by Hatsuda-Lee,
we briefly recall QSR for the vector meson mass in the vacuum.
The correlation function in the vacuum defined in (\ref{eq1.6})
has the structure
\begin{eqnarray}
\Pi_{\mu\nu}^0(q)= (q_\mu q_\nu -g_{\mu\nu}q^2)\Pi_1(q^2).
\label{eq3.4}
\end{eqnarray}
In QSR one starts with the dispersion relation for $\Pi_1(q)$
(See Appendix C):
\begin{eqnarray}
\Pi_1(q^2) = {q^2 \over \pi}\int_{0+}^\infty\,d\,s { 
{\rm Im}\Pi_1(s)
\over s(s -q^2)} + \Pi_1(0),
\label{eq3.5}
\end{eqnarray}
where we introduced 
one subtraction to avoid the logarithmic divergence.
In the deep Euclidean region $q^2\to -\infty$, 
$\Pi_1(q^2)$ in the the left hand side of 
(\ref{eq3.5})
has the OPE expression including the operators up to
dim.=6 as
\begin{eqnarray}
\Pi_1^{\rm OPE}(q^2) = -{ 1 \over 8\pi^2 } 
{\rm ln}(-q^2) + { A \over q^4}
+ {B \over q^6},
\label{eq3.6}
\end{eqnarray}
where $A$ and $B$ are respectively the 
sums of dim.=4 and dim.=6
condensates, and the 
perturbative correction factor $1 + {\alpha_s \over \pi}$
to the first term is omitted for simplicity.
We also suppressed the scale dependence in each 
term in (\ref{eq3.6}). 
The spectral function in 
(\ref{eq3.5}) is often modeled by the sum of the pole
contribution from the vector meson and the continuum 
contribution:
\begin{eqnarray}
{1 \over \pi}{\rm Im}\Pi_1(s)
=F'\delta(s-m_V^2)+{1\over 8\pi^2}\theta(s-s_0),
\label{eq3.7}
\end{eqnarray}
where $F'=f_V^2 m_V^2$.
With this form in (\ref{eq3.5}) together with (\ref{eq3.6}), 
one gets the
sum rule relation (See Appendix C) as
\begin{eqnarray}
{1 \over 8\pi^2}{\rm ln}\left( {s_0-q^2 \over -q^2}\right)
+{ A \over q^4} + { B \over q^6} = {F' \over m_V^2 - q^2}.
\label{eq3.8}
\end{eqnarray}
Hatsuda-Lee considered the sum rule
for $\Pi_1^{\rm NM}(\omega^2)= 
\Pi^{{\rm NM}\mu}_\mu (\omega,\mbox{\boldmath $q$}=0)/(-3\omega^2)$.
The QSR for $\Pi_1^{\rm NM}(q^2)$ is reduced to 
(\ref{eq3.8}) at $\rho_N\to 0$
limit.  At $\mbox{\boldmath $q$}=0$, $\Pi_1^{\rm NM}(q^2)$ becomes
\begin{eqnarray}
\Pi_1^{\rm NM}(q^2)=\Pi_1(q^2)+{\rho_N\over 2M_N}{ T(q^2)\over q^2}
+O(\rho_N^{5/3}).
\label{eq3.9}
\end{eqnarray}
Thus one has to analyze $T(q^2)/q^2$ to understand the density
dependence in $\Pi_1^{\rm NM}(\omega^2)$.
We write the
dispersion relation for $T(q^2)/q^2$: 
\begin{eqnarray}
{T(q^2) \over q^2} = \int_{0+}^\infty\,d\,s { \rho(s) \over
s(s-q^2)} + {T(0) \over q^2},
\label{eq3.10}
\end{eqnarray}
where the pole contribution 
at $q^2=0$ is explicitly taken care of by $T(0)$.
Substituting the spectral function (\ref{eq2.6})
in this equation and equating it to the OPE side, one gets
the QSR relation for the case (ii) as
\begin{eqnarray}
{ \alpha \over q^4} + {\beta \over q^6}
={ a' \over (m_V^2 - q^2)^2} + { b' \over m_V^2 -q^2}
+{(T^{\rm Born}(0)-b') \over s_0 -q^2} + { T^{\rm Born}(0) \over q^2},
\label{eq3.11}
\end{eqnarray}
where
\begin{eqnarray}
a'= {a \over m_V^2},\ \ \ \ \ b'={a \over m_V^4}+{b \over m_V^2},
\label{eq3.12}
\end{eqnarray}
and the relation $T(0)=T^{\rm Born}(0)$ 
is used in the last term of 
(\ref{eq3.11}).
We note that
(\ref{eq3.11}) is nothing but the relation obtained by
dividing both sides of (\ref{QSR}) by $q^2$ for the case (ii),
which guarantees the absence of the $1\over q^2$ term
in the right hand side.
(Note that the condition $T(0)=T^{\rm Born}(0)$ itself is 
not required
to guarantee this consistency condition.)
Using (\ref{eq3.8}) and (\ref{eq3.11}) in (\ref{eq3.9}), 
we can construct
the QSR for $\Pi_1^{\rm NM}(q)$ in the linear 
density approximation:
\begin{eqnarray}
{1 \over 8\pi^2}{\rm ln}\left( {s_0 - q^2 \over -q^2} \right)
+{ A+\widetilde{\alpha} \over q^4}+{ B+ \widetilde{\beta} 
\over q^6}
={ F' + \widetilde{b'} \over m_V^2 - q^2} +
{ \widetilde{a'} \over (m_V^2-q^2)^2}+ 
{ \widetilde{T}^{\rm Born}(0) -\widetilde{b'}
\over s_0 -q^2} + { \widetilde{T}^{\rm Born}(0) \over q^2},
\nonumber\\
\label{eq3.13}
\end{eqnarray}
where $\widetilde{\alpha}={\rho_N \over 2M_N}\alpha$,
$\widetilde{a'}={\rho_N \over 2M_N}a'$, etc.  
To $O(\rho_N)$ accuracy
(\ref{eq3.13}) can be rewritten as
\begin{eqnarray}
\frac{1}{8\pi^2}\mbox{ln}\left(\frac{s_0^*-q^2}{-q^2}\right) 
+ \frac{A^*}{q^4} + \frac{B^*}{q^6} = 
\frac{F'^{*}}{m_{V}^{*2}-q^{2}} 
+\frac{\widetilde{T}^{\rm Born}(0)}{q^{2}},
\label{eq3.14}
\end{eqnarray}
with
\begin{eqnarray}
A^{*} = A + \widetilde{\alpha}, \ \ \ \  
B^{*} = B +\widetilde{\beta},
\label{eq3.15a}
\end{eqnarray}
\begin{eqnarray}
F'^* =F'+\widetilde{b'},\ \ \ \
m_V^{*2}=m_V^2 - { \widetilde{a'} \over F'},
\ \ \ \  
s_{0}^{*} = s_{0} - 8\pi^2( \widetilde{T}^{\rm Born}(0) 
- \widetilde{b'}).
\label{eq3.15b}
\end{eqnarray}
From the above demonstration, it is now clear that
our analysis of $T_{\mu\nu}$ in the previous section
(case (ii)) leads to (\ref{eq1.9}) for $\Pi_1^{\rm NM}$
by the identification $\rho_{sc}=\widetilde{T}^{\rm Born}(0)$.
In fact $\rho_{sc}={-\rho_N \over 2M_N}$ for 
the $\rho$, $\omega$ mesons
and $\rho_{sc}=0$ for the $\phi$ meson in \cite{HL}, which is 
consistent with (\ref{eq2.9}).
The mass shift in (\ref{eq3.15b}) is obviously the same as
given in (\ref{eq3.2}).
We should emphasize that it is our constraint relation
$T^{\rm ph}(0)=T^{\rm Born}(0)$ in the analysis of the
scattering lengths
which leads to the same sum rule for $\Pi_1^{\rm NM}(q^2)$
as in \cite{HL}.
Our use of low energy theorem is in parallel with
the calculation of the Landau damping term $\rho_{sc}$ 
from the Born diagram in \cite{HL}.
If one did not have such information
on $T(0)$, one would have to use the approach in \cite{Koike}
with the matrix elements of the dim.=8 or higher operators.

We point out, however, a small difference from \cite{HL}.
From the first and the third relation in (\ref{eq3.15b}), one obtains
\begin{eqnarray}
F'^* - F' = {1 \over 8\pi^2}\left( s_0^* - s_0 \right) + 
\widetilde{T}^{\rm Born}(0),
\label{eq3.15p}
\end{eqnarray}
which is the same as the first FESR relation obtained 
from $\Pi_1^{\rm NM}$. (In FESR our 
present analysis is completely equivalent
to \cite{HL}.)
Namely the shift of $F'$ is determined by that of $s_0$.  
In the Borel sum rule in \cite{HL}, $F'^*$ and $s_0^*$
are regarded as independent fitting parameters.
But if one recognizes that the QSR
for $T_{\mu\nu}$ is independent from
that for $\Pi_{\mu\nu}$,
it is easy to see that this condition has to be also satisfied
in the approach of \cite{HL}.  In fact, in
(\ref{eq1.11}) which was derived purely mathematically 
from the sum rule
in \cite{HL},
absence of $1/q^2$ term in the left hand side 
of (\ref{eq1.11}) imposes the consistency requirement
in the right hand side of (\ref{eq1.11}), which is exactly
(\ref{eq3.15p}).  
Since HL took the view that $\rho_{sc}$ is calculable
(owing to the low energy theorem),
they could have eliminated $\delta F'$ or $\delta s_0$  
from the outset.
In our BSR for $T_{\mu\nu}$, we were lead to use the condition
(\ref{eq3.15p}) explicitly, which is imposed by the low energy theorem
$T(0)= T^{\rm Born}(0)$.
In our opinion, this is more natural because
the QSR for $T_{\mu\nu}$ is completely independent from
the one for $\Pi^0_{\mu\nu}$, i.e.,
the density $\rho_N$ is simply 
an external parameter 
which connects these 
quantities in the sum rule for $\Pi^{\rm NM}_1$.

Although the mass shifts discussed in this section 
are essentially the same
as those in \cite{HL}, the numerical values in 
(\ref{table2}) are approximately factor two
smaller
than those in \cite{HL}, especially for the $\rho$ and $\omega$ mesons.
This is mainly because their calculation is done
at the chiral limit
(they ignored a correction due to the condensate 
$m_q\langle \bar{\psi}\psi\rangle$), and correspondingly their
value for the continuum 
threshold $s_0$
is different from ours.  They used $s_0=1.43$ GeV$^2$ 
for $\rho$, $\omega$ in the vacuum.  
Another reason is that their QSR
was for the total sum of $\Pi^0_{\mu\nu}$ and $T_{\mu\nu}$,
the latter being small ($O(\rho_N)$) correction to the 
former as noted above, while our QSR is for the latter.
These differences eventually leads to 
factor-two difference in the mass shifts 
at around nuclear matter density. 
Hatsuda claims\,\cite{Hatsuda} that, although
$m_V^*/m_V$ at $\rho_N=0$ in \cite{HL,JL} 
is consistent with our scattering lengths, 
the mass shift in \cite{HL,JL} at higher $\rho_N$ 
becomes bigger than expected from the scattering length,
with the reasoning that the scattering length
can be used only at very close to zero density
and the prediction in \cite{HL} contains more than that.
This deviation, however, should not be regarded as 
a meanigful one, since
the OPE side includes only $O(\rho_N)$ density effect
and therefore only the $O(\rho_N)$ effect
represented by the scattering length
is a valid physical prediction.

It is probably useful to add a brief comment on the 
calculation
of $\rho_{sc}$ in \cite{HL}.
Using the general relation
\begin{eqnarray}
\lim_{\mbox{\boldmath $q$}\to 0}\Pi_1^{\rm NM}(\omega, 
\mbox{\boldmath $q$})
=\lim_{\mbox{\boldmath $q$}\to 0}
{\Pi_{00}^{\rm NM}(\omega,\mbox{\boldmath $q$})
\over|\mbox{\boldmath $q$}|^2},
\label{eq3.16}
\end{eqnarray}
they calculated $\rho_{sc}$ from the spectral function
of $\Pi_{00}^{\rm NM}(\omega,
\mbox{\boldmath $q$})/|\mbox{\boldmath $q$}|^2$
which corresponds to 
$T_{00}(\omega,
\mbox{\boldmath $q$})/|\mbox{\boldmath $q$}|^2$ in our method.  
They included 
the pole contributions which appear at $\omega=\pm 0$ and
ignored the contributions from $\omega=\pm 2M_N$.
But this treatment suffices as long as one needs the value of 
$T^{\rm Born}(0)$.
The residue at $\omega=\pm 0$ (=$-1/2$ for $\rho$, $\omega$ meson)
of $\lim_{\mbox{\boldmath $q$}\to 0}T_{00}
(\omega,\mbox{\boldmath $q$})/|\mbox{\boldmath $q$}|^2$
precisely gives $T^{\rm Born}(0)$.  (See Appendix B.)
As was noticed below (\ref{QSR}), their neglection
of the poles at $\omega=\pm 2M_N$ in 
$\lim_{\mbox{\boldmath $q$}\to 0}
\Pi_{00}^{\rm NM}(\omega,
\mbox{\boldmath $q$})/|\mbox{\boldmath $q$}|^2$
corresponds to
the assumption
that those contributions are taken care of
in the continuum part of (\ref{eq2.10}) 
( $1/(s_0-q^2)$ term) in our language.

\section{Comments and Replies to Hatsuda-Lee-Shiomi (HLS)}
\setcounter{equation}{0}
\renewcommand{\theequation}{\arabic{section}.\arabic{equation}}

It is by now clear that our present QSR analysis 
on the current-nucleon 
forward scattering amplitude is essentially equivalent 
to the medium QSR for the vector mesons in \cite{HL}.
Namely their result is certainly interpretable in terms
of the $V-N$ scattering lengths.
Although this already resolves the essential controversy
between \cite{Koike} and \cite{HL}, we summarize in the following
our replies and comments to HLS\,\cite{HLS}.

\vspace{1cm}
\noindent
(1) HLS claims that $V-N$ scattering lengths
are not calculable in QSR without including dim.=8 matrix elements.
This is because phenomenological side contains 
three unknown parameters
but finite energy sum rules (FESR) provide only two relations.
(Sec. III.B of \cite{HLS})  

\vspace{0.5cm}
\noindent
Reply:

In our present analysis, we eliminated one parameter
by the constraint relation
at $q^\mu =0$ and thus have two unknown parameters to be determined
by QSR.  
This constraint relation due to the low energy theorem
renders our analysis 
equivalent to HL in the FESR.  

\vspace{1.0cm}

\noindent
(2) HLS claims $\Pi^{\rm NM}
(\omega^2)=\omega^2\Pi_1^{\rm NM}(\omega^2)$
is not usable to predict the mass of the vector mesons 
either in medium or in the vacuum.
(Sec. III.C of \cite{HLS})  

\vspace{0.5cm}
\noindent
Reply:

This argumentation is based on the number of 
available FESRs' and the 
Borel stability of
the sum rules.  As is shown in Appendix C, the QSR itself (before
Borel transform) is the same for 
$\Pi^{\rm NM}(\omega^2)$ and $\Pi_1^{\rm NM}(\omega^2)$
as long as one starts with the same consistent assumption
for these quantities.  
The QSR for $\Pi^{\rm NM}$ is simply the one obtained
by multiplying $\omega^2$ to $\Pi_1^{\rm NM}$.
Accordingly the FESRs' are the same.
Whether one applies Borel transform before or 
after multiplying $\omega^2$
to $\Pi_1^{\rm NM}(\omega^2)$ causes numerical difference,
especially because one loses information from the
polynomial terms in the BSR method.  
We agree that applying Borel transform
to $\Pi_1^{\rm NM}$ leads to more 
stable Borel curve than applying to $\Pi^{\rm NM}$.
We, however, note that the reason HL obtained
the stable Borel curve for $\Pi_1^{\rm NM}$ is that
the Borel curve for $\Pi_1$ in the vacuum is very stable 
and the curve for ${\rho_N\over 2 M_N} {T(q^2)\over q^2}$
(see (\ref{eq3.9})) is only an $O(\rho_N)$ correction to the former.
In our case, what is plotted in Figs. 2 and 3 are the
Borel curves for $T(q^2)$ itself.

The authors of \cite{JL} raised a similar criticism against
\cite{Koike} and claimed that they have clarified the 
origin of discrepancy
between \cite{HL} and \cite{Koike}.  
But this does not solve the problem.

In \cite{Hatsuda}, it was advertised that
the Borel curves for $m_V^*$ in \cite{HL}
are more stable than 
those for our scattering lengths shown in Figs. 2 and 3.
But the reason for this is obvious.  The stability of the Borel curve
for $m_V$ is excellent in the vacuum, and the density
effect based on the scattering length is simply a small
$O(\rho_N)$ correction to it.

\vspace{1.0cm}
\noindent
(3) HLS claims that the $V-N$ scattering lengths and the mass 
shift of the vector mesons in the nuclear matter have no direct
relation due to the momentum dependence of the $V-N$ forward 
scattering amplitude.  
They also claim that
the analysis in \cite{HL} did not use this relation.
(Sec. III.A of \cite{HLS})

\vspace{0.5cm}
\noindent
Reply: 

As is noted in the introduction, the analysis in \cite{HL} is
{\it mathematically} 
equivalent to the QSR for $T_{\mu\nu}$ shown in
(\ref{eq1.11}).  
The right hand side of (\ref{eq1.11}) is precisely
reproduced by the spectral function 
shown in (\ref{eq2.6}) and the Born
contribution to $T_{\mu\nu}$.  Thus the physical effect
which caused the mass shift in \cite{HL} is essentially
the same as the one based on the $V-N$ scattering lengths.

HLS stressed the 
importance of the momentum dependence of $T_{\mu\nu}$.  However,
it is not conspicuous in the OPE side.  
So one can not claim its importance
in the phenomenological side from the QSR analysis itself.
In fact the effect of the fermi motion of 
the nucleon can be included
in the OPE side, but they are at least $O(\rho_N^{5/3})$ and
they can be neglected as was shown in sec. IV of \cite{HLS}.
How come one can claim the importance 
of the effect which is negligible
in the OPE side?   
Since the common starting point of our analysis was
the linear density approximation 
to the OPE side shown in (\ref{eq1.4})
the negligible effect in (\ref{eq1.4}) should be taken
as the effect which is either 
negligible in the phenomenological side or
beyond the resolution of the analysis.

The phenomenological basis on 
which HLS emphasize the effect of Fermi motion
of the nucleons is as follows:
Nucleon's fermi momentum is
$p_f=270$ MeV in Nuclear matter and thus one should take into
account the $\rho-N$
scattering from $\sqrt{s}=m_{\rho}+M_N=1709$ MeV 
through $\sqrt{s}=[(m_\rho
+\sqrt{M_N^2+p_f^2})^2-p_f^2]^{1/2}=1726$
MeV.  In this interval there are
some $s$-channel resonances such as $N(1710)$ and $N(1720)$
which couple to $\rho-N$ channel, thus 
$T_{\mu\nu}$ should change rapidly in this
interval.  However, these resonances 
together with the other near resonances
($N(1700)$, $\Delta(1700)$) have broad widths of over $100$ MeV
and that 
the whole interval 
$0<|\mbox{\boldmath $p$}| <p_f$ is buried under these
broad resonance regions.
In this situation,
it is unlikely that the $V-N$ phase shift changes rapidly
in this interval.  
It might be a good approximation to take the $T$-matrix
at $\mbox{\boldmath $p$}=0$ as a representative value of it.  
How about $\phi$ meson?  The $\phi-N$ scattering occurs from
$\sqrt{s}=m_{\phi}+M_N=1960$ MeV 
through $\sqrt{s}
=[(m_\phi+\sqrt{M_N^2+p_f^2})^2-p_f^2]^{1/2}=1980$
MeV.  In this interval there is no resonance which could couple to
$\phi-N$ system.  The situation is better.

In \cite{FH,KM2}, there is a debate on the interpretation
of the nucleon sum rule in the nuclear medium.
We agree with the interpretation of \cite{KM2}.
A difference between the sum rules for the $V-N$ and $N-N$ 
interactions is the smallness of
the obtained $V-N$ scattering lengths, which, together with
the argument above, may
justify the use of (\ref{eq1.8}) to predict the mass
shift in the linear density approximation.

One can organize the finite temperature ($T$) QSR in a similar way,
replacing the Fermi gas of nucleons by the ideal gas of pions
\,\cite{HKL,Koi}.  In this formalism, the $T$-dependence
of correlation functions 
comes from the current-pion forward amplitude.
Since the pion-hadron 
scattering lengths are zero in the chiral limit,
there is no $O(T^2)$ mass shift\,\cite{Koi,EI}.  
This is in parallel with
our present analysis that $O(\rho_N)$-dependence of
the mass is determined by the scattering length.

\section{Summary and Conclusions}
\setcounter{equation}{0}
\renewcommand{\theequation}{\arabic{section}.\arabic{equation}}

In this paper, 
we have presented a new analysis on the $\rho$, $\omega$ 
and $\phi$ meson-nucleon spin-isospin 
averaged scattering lengths
$a_V$ ($V=\rho,\omega,\phi$)
in the framework of the QCD sum rule.  Essential difference from
the previous calculation in \cite{Koike} is that
the parameters in the spectral 
function of the vector current-nucleon 
forward amplitude is constrained by the relation at $q^\mu =0$
(low energy theorem for the vector-current nucleon scattering amplitude).
We obtained small negative values for $a_V$ as
\begin{eqnarray}
a_{\rho} &=& -0.47\pm 0.05\ \mbox{fm}, \nonumber \\
a_{\omega} &=& -0.41 \pm 0.05\ \mbox{fm}, \nonumber \\
a_{\phi} &=& -0.15\pm 0.02\ \mbox{fm}.
\end{eqnarray}
This suggests that these $V-N$ interactions are weakly attractive
in contrast to the previous study \cite{Koike}.
Since the form of the spectral function is greatly simplified,
these numbers should be taken as a rough estimate of the 
order of magnitude.
In the axial vector channel, the method works as a tool to
introduce a correction to the current algebra calculation
due to the higher dimensional operators.  Present application
to the vector channel in which current algebra technique
does not work is suggestive in that the QSR provides us with a 
possibility to express the $V-N$ scattering lengths in terms of
various nucleon matrix elements.

If one applies above $a_V$s' to the
vector meson masses in the nuclear medium in the linear
density approximation, one gets for the mass shifts as 
\begin{eqnarray}
\delta m_{\rho} &=& -45 \sim -55\ \mbox{MeV}\  
(6 \sim 7\%), \nonumber \\
\delta m_{\omega} &=& -40 \sim -50\ \mbox{MeV}\  
(5 \sim 6\%), \nonumber \\
\delta m_{\phi} &=& -10 \sim -20\ \mbox{MeV}\  (1 \sim 2\%),
\end{eqnarray}
at the nuclear matter density.  
We have shown that the physical
content of the mass shifts 
discussed in \cite{HL} 
are essentially the one due to the scattering
lengths shown above and have resolved the discrepancy
between \cite{HL} and \cite{Koike}. 

One might naturally ask whether the previous QSR\,\cite{Koike} 
for the scattering lengths is wrong or not.
Compared with \cite{Koike}, the present 
analysis utilizes more available information, i.e. the constraint
from the low energy theorem.  In this sence, 
one may say that the present analysis is a more
sound one.
If one did not have such information
on $T(0)$, one would have to use the approach in \cite{Koike}
with the inclusion of the 
matrix elements of the dim.=8 or higher operators.
In this sence, the way of constructing sum rule itself
in \cite{Koike} is also correct. 

Another point we wish to emphasize is that
regardless of the availabilty of the information
on $T(0)$ (such as the low energy theorem),
the sum rule for $m_V^*$ in (\ref{eq1.9})\,\cite{HL} 
in the linear density approximation is automatically
equivalent to the mass shift due to the scattering lengths
as is shown in (\ref{eq1.11}).

Finally, we wish to make some comments on 
the interpretation in the literature
about the mass shifts of the vector mesons in the nuclear medium.
Several effective theories for the vector mesons 
($\rho$, $\omega$)\,\cite{SMS} 
predicts decreasing masses in the nuclear medium, and 
the magnitude of the mass shifts is quite similar to the 
QSR analysis in \cite{HL}.
Accordingly, the ``similarity'' and ``consistency'' 
between QSR in medium and 
the effective theories has been erroneously advertized
in the literature\,\cite{Hatsuda, HLS}.
The essential ingredient of the mass shifts
predicted by those effective theories
is the polarization in the Dirac sea of the nuclear
medium, which leads to a smaller effective mass of the 
nucleon in the nuclear medium.  
If one switch off this effect, the vector meson propagators
receives only the effects of the Fermi sea of the nucleons,
which leads to small positive mass shifts of those vector 
mesons\,\cite{Chin}.  One has to recognize that
the physical effect which QSR for the vector mesons in medium
is enjoying is simply the scattering with this Fermi sea 
of the nucleons (through the forward scattering amplitude
with the nucleon) which has the same mass as in the vacuum,
and accordingly the QSR in medium does not pick up any effect
of the polarization of the Dirac sea of the nucleons.
Similarity in prediction on the mass shift between the medium 
QSR\,\cite{HL}  and 
the effective theories\,\cite{SMS} looks fortuitous and
rather causes new problems.
As has been clarified in this work, the medium QSR
presented by \cite{HL} should be interpreted as
a QCD sum rule analysis on the vector current-nucleon forward amplitude,
and should not be interpreted as a method which picks up
an effect of the vacuum polarization due the finite baryon number
density.   It is misleading to celebrate the medium QSR
in \cite{HL} as
a tool to incorporate the 
effect of ``change of QCD vacuum''
due to the finite baryon density.

\vskip 0.5cm

\centerline{\bf Acknowledgement}

We thank O. Morimatsu for useful comments on the manuscript.

\newpage

\appendix
\renewcommand{\theequation}{\Alph{section}.\arabic{equation}}
\renewcommand{\thesection}{\Alph{section}.}
\centerline{\large{\bf APPENDIX}}

\section{Spin-average of $T_{\mu\nu}$}
\setcounter{equation}{0}
$T_{\mu\nu}$ in (\ref{eq2.1}) can be decomposed as
\begin{eqnarray}
T_{\mu\nu}= - \left( g_{\mu\nu} 
- { q_\mu q_\nu \over q^2} \right)T_1
 + { 1 \over M_N^2}\left( p_\mu 
- { p\cdot q \over q^2}q_\mu \right)
\left( p_\nu - 
{ p\cdot q \over q^2}q_\nu \right) T_2 +\cdot\cdot\cdot,
\label{a1}
\end{eqnarray}
where $\cdot\cdot\cdot$ denotes the pieces which depend on the
nucleon-spin and vanish after averaging over the nucleon spin.
Averaging over the helicities of the vector current can be done
by the following procedure\,\cite{err}:
\begin{eqnarray}
T(\omega,\mbox{\boldmath $q$})
\equiv {1\over 3}\sum_{\epsilon\cdot q =0}\epsilon^{(h)}_\mu (q) 
T_{\mu\nu}\epsilon_\nu^{*(h)}(q)
={1\over 3}\left( -g_{\mu\nu} 
+ {q_\mu q_\nu \over q^2 }\right)T^{\mu\nu}
={-1 \over 3}T_\mu^\mu
=T_1 -{1\over 3}\left( 1 - 
{ (p\cdot q)^2 \over M_N^2 q^2 } \right) T_2,
\nonumber\\
\label{a2}
\end{eqnarray}
where the use has been made of the relation $\sum_{\epsilon\cdot q=0}
\epsilon_\mu^{(h)}\epsilon_\nu^{*(h)}
=-g_{\mu\nu} + { q_\mu q_\nu \over q^2}$.
By this spin-averaging $T$-matrix is 
projected to the combination\,\cite{err},
\begin{eqnarray}
& &{1\over 2}{1 \over 3}
\left( {\cal T}_{1{1\over 2},1{1\over 2}}+
{\cal T}_{1{-1\over 2},1{-1\over 2}}
+{\cal T}_{-1{1\over 2},-1{1\over 2}}
+{\cal T}_{-1{-1\over 2},-1{-1\over 2}}
+{\cal T}_{0{1\over 2},0{1\over 2}}
+{\cal T}_{0{-1\over 2},0{-1\over 2}} \right)\nonumber\\
&=&{1 \over 6}\left( \sum_{J_z=\pm 1/2,\pm 3/2}
{\cal T}(J=3/2,J_z)
+\sum_{J_z=\pm 1/2}{\cal T}(J=1/2,J_z) \right)
\nonumber\\
&\stackrel{\omega\to m_V,\mbox{\boldmath $q$}\to 0}
{\longrightarrow}& 
{1 \over 6}8\pi(m_V + M_N)(4a_{3/2}+2a_{1/2})
\nonumber\\
&=& 8\pi(m_V+M_N)a_V
\end{eqnarray}
with the $V-N$ spin-averaged scattering length 
$a_V={1\over 3}(2a_{3/2}+a_{1/2})$.

\section{Born diagram contribution to $T_{\mu\nu}$}
\setcounter{equation}{0}

Here we
summarize the Born diagram contribution (Fig. 1) to
$T_{\mu\nu}$.  
This is the only contribution to the vector current-nucleon
forward amplitude in the $q^\mu \to 0$ limit.
For $J_\mu^{\rho,\omega}$, it becomes after averaging over
the nucleon spin,
\begin{eqnarray}
T_{\mu\nu}(p,q)
&=&{-1 \over (p+q)^2 - M_N^2 +i\varepsilon} 
{1 \over 4}{1 \over 2}{\rm Tr}\left[ 
\gamma_\mu (\rlap/{\mkern-1mu p} 
+\rlap/{\mkern-1mu q} 
+ M_N)\gamma_\nu (\rlap/{\mkern-1mu p} + M_N)\right]
\nonumber\\
&+& { -1 \over (p-q)^2 -M_N^2 +i\varepsilon}
{1 \over 4}{1 \over 2}{\rm Tr}\left[ 
\gamma_\mu (\rlap/{\mkern-1mu p} 
-\rlap/{\mkern-1mu q} + M_N)\gamma_\nu 
(\rlap/{\mkern-1mu p} + M_N)\right].
\label{b1}
\end{eqnarray}
In the following we consider 
the two quantities, $T=T_{\mu}^\mu/(-3)$ and 
$T_{00}$.  A straightforward calculation gives
\begin{eqnarray}
T(p,q)={1\over 3 }
\left[ {M_N^2 - p\cdot q \over q^2 + 2 p\cdot q+ i \varepsilon}+
{M_N^2 + p\cdot q \over q^2 - 2 p\cdot q+ i\varepsilon} \right].
\label{b2}
\end{eqnarray}
At $\mbox{\boldmath $p$}=0$ and $\mbox{\boldmath $q$}=0$, $T$ becomes
\begin{eqnarray}
T(\omega, \mbox{\boldmath $q$}=0)= { 2M_N^2 \over \omega^2 - 4 M_N^2}.
\label{b3}
\end{eqnarray}
On the other hand, 
$T_{00}$ at $\mbox{\boldmath $p$}=0$ is calculated to be 
\begin{eqnarray}
T_{00}(\omega,\mbox{\boldmath $q$})
={ 2M_N^2 \mbox{\boldmath $q$}^2 \over (q^2)^2 - 4 M_N^2\omega^2}.
\label{b3p}
\end{eqnarray} 
From (\ref{b3}) and (\ref{b3p}), one sees
\begin{eqnarray}
\lim_{|\mbox{\boldmath $q$}|\to 0}
{ T_{00}(\omega, \mbox{\boldmath $q$}) 
\over |\mbox{\boldmath $q$}|^2 }
= \lim_{|\mbox{\boldmath $q$}|\to 0}
{T_\mu^\mu(\omega,\mbox{\boldmath $q$}) \over -3\omega^2}
={2M_N^2 \over \omega^2 (\omega^2 -4M_N^2)}.
\label{b4}
\end{eqnarray}
To understand the relation between the Landau damping term
$\rho_{sc}$ in \cite{HL} and $T^{\rm Born}(0)$, 
we consider the spectral function
\begin{eqnarray}
\rho_{00}(\omega,\mbox{\boldmath $q$})&=&{1 \over \pi}
\left(\theta(\omega)-\theta(-\omega)\right)
{\rm Im}T_{00}(\omega,\mbox{\boldmath $q$})
\label{b5}\\
&\equiv &\rho^{(0)}_{00}(\omega,\mbox{\boldmath $q$})
+\rho^{(1)}_{00}(\omega,\mbox{\boldmath $q$}),
\label{b6}
\end{eqnarray}
where
\begin{eqnarray}
\rho^{(0)}_{00}(\omega,\mbox{\boldmath $q$})&=&
{M_N(2M_N+\omega) \over \sqrt{M_N^2+\mbox{\boldmath $q$}^2}}
\delta(\omega + M_N-\sqrt{M_N^2 + \mbox{\boldmath $q$}^2})
\nonumber\\
&-&{M_N(2M_N-\omega) \over \sqrt{M_N^2+\mbox{\boldmath $q$}^2}}
\delta(\omega - M_N+\sqrt{M_N^2 + \mbox{\boldmath $q$}^2}),
\label{b7}\\
\rho^{(1)}_{00}(\omega,\mbox{\boldmath $q$})&=&
-{M_N(2M_N+\omega) \over \sqrt{M_N^2+\mbox{\boldmath $q$}^2}}
\delta(\omega + M_N+\sqrt{M_N^2 + \mbox{\boldmath $q$}^2})
\nonumber\\
&+&{M_N(2M_N-\omega) \over \sqrt{M_N^2+\mbox{\boldmath $q$}^2}}
\delta(\omega - M_N-\sqrt{M_N^2 + \mbox{\boldmath $q$}^2}).
\label{b8}
\end{eqnarray}
At $\mbox{\boldmath $q$}\to 0$, 
$\rho^{(0)}_{00}(\omega,\mbox{\boldmath $q$})$ has a pole at
$\omega=\pm 0$ and $\rho^{(1)}_{00}(\omega,
\mbox{\boldmath $q$})$ has poles at
$\omega=\pm 2M_N$.
If we define $T_{00}^{(0,1)}(\omega,
\mbox{\boldmath $q$})$ by the dispersion integral 
\begin{eqnarray}
T_{00}^{(0,1)}(\omega,\mbox{\boldmath $q$})= 
\int_{-\infty}^{\infty}\,d\,u 
{\rho_{00}^{(0,1)}(u,\mbox{\boldmath $q$}) 
\over u-\omega - i\varepsilon },
\label{b9}
\end{eqnarray}
then we get
\begin{eqnarray}
\lim_{\mbox{\boldmath $q$}\to 0}{ T_{00}^{(0)}(\omega, 
\mbox{\boldmath $q$}) \over |\mbox{\boldmath $q$}|^2}
={-1 \over 2\omega^2},
\label{b10}\\
\lim_{\mbox{\boldmath $q$}\to 0}{ T_{00}^{(1)}(\omega, 
\mbox{\boldmath $q$}) \over |\mbox{\boldmath $q$}|^2}
={1 \over 2(\omega^2 - 4M_N^2)}.
\label{b11}
\end{eqnarray}
Thus
$\lim_{\mbox{\boldmath $q$}\to 0}T_\mu^\mu/(-3\omega^2)$, 
$\lim_{\mbox{\boldmath $q$}\to 0}T^{(0)}_{00}(\omega,
\mbox{\boldmath $q$})/|\mbox{\boldmath $q$}|^2$ and
$\lim_{\mbox{\boldmath $q$}\to 0}T_{00}(\omega,
\mbox{\boldmath $q$})/|\mbox{\boldmath $q$}|^2$
have a pole at $\omega=0$ with the same residue $-1/2$.
In the QSR in the medium\,\cite{HL}, 
HL essentially calculated
$\lim_{\mbox{\boldmath $q$}\to 0}T^{(0)}_{00}
(\omega,\mbox{\boldmath $q$})/|\mbox{\boldmath $q$}|^2$
starting from the spectral function for the Landau damping term.

\section{Dispersion relations}
\setcounter{equation}{0}

Here we summarize the basics of the dispersion relations,
since they are essential in carrying out the QSR analyses.
The contents in this appendix is trivial, as long as 
one is careful
enough.  We dare to add this appendix, since some of the
criticisms raised in \cite{HL} appear to originate from the
misunderstanding of the dispersion relation.

In the text, scalar functions $\Pi(q^2)$ and $\Pi_1(q^2)$ are
defined from the vector current correlator in the vacuum as
\begin{eqnarray}
\Pi^0_{\mu\nu}(q)=(q_\mu q_\nu -g_{\mu\nu}q^2)\Pi_1(q^2)
=\left( { q_\mu q_\mu \over q^2}-g_{\mu\nu}\right)\Pi(q^2).
\label{eqc.1}
\end{eqnarray}
In the medium, $\Pi^{\rm NM}(q^2)$ and $\Pi_1^{\rm NM}(q^2)$
are defined at $\mbox{\boldmath $q$}=0$ as
\begin{eqnarray}
\Pi^{\rm NM}(\omega^2)
=\Pi^{{\rm NM}\mu}_\mu(\omega,0)/(-3), \ \ \ \ \ 
\Pi^{\rm NM}_1(\omega^2)=\Pi^{\rm NM}(\omega^2)/\omega^2.
\label{eqc.2}
\end{eqnarray}
$\Pi^{\rm NM}\rightarrow \Pi$ and $\Pi_1^{\rm NM}\rightarrow 
\Pi_1$
as $\rho_N \to 0$.
In the following, we discuss the dispersion
relations for $\Pi^{\rm NM}_1$ and
$\Pi^{\rm NM}$.  (Those for $\Pi_1$ and $\Pi$ are parallel.)
$\Pi_1^{\rm NM}(\omega^2)$ has an
isolated pole of first order at $\omega^2=0$
and has a Laurent expansion as $\Pi_1^{\rm NM}
(\omega^2)={c_{-1}\over \omega^2}
+c_0+c_1\omega^2+\cdots$.  Anticipating a logarithmic divergence
in the dispersion integral, 
we introduce one subtraction for $\Pi_1^{\rm NM}$.
Applying the Cauchy's theorem, one gets
\begin{eqnarray}
{1 \over 2\pi i}\oint_{C_1}\,ds\, { \Pi_1^{\rm NM}(s) \over
s(s-\omega^2)}&=&{\Pi^{\rm NM}_1(\omega^2) \over \omega^2}+
\lim_{s\to 0}{d\over ds}\left[ 
{s\Pi_1^{\rm NM}(s)\over s-\omega^2}\right]
\nonumber\\
&=& {\Pi_1(\omega^2) \over \omega^2} - {c_0\over \omega^2}
-{c_{-1}\over \omega^4},
\label{eqc.3}
\end{eqnarray}
where the closed path $C_1$ is taken as shown in Fig. 5.
Equation (\ref{eqc.3}) can be written as
\begin{eqnarray}
\Pi_1^{\rm NM}(\omega^2)={c_{-1}\over\omega^2}
+c_0+{\omega^2\over\pi}
\int_{0+}^\infty \,ds\,{{\rm Im}\Pi_1^{\rm NM}(s)
\over (s-i\varepsilon)
(s-\omega^2-i\varepsilon)}.
\label{eqc.4}
\end{eqnarray}
In (\ref{eqc.4}) the pole contribution at $\omega^2=0$
is explicitly taken into account separately and thus the integral
along the positive real axis excludes $s=0$.  This is also the same 
for (\ref{eqc.5}) below.
For $\Pi^{\rm NM}$, we need additional subtraction.  
We thus consider
\begin{eqnarray}
{1\over 2\pi i}\oint_{C_1}\,ds\,{\Pi^{\rm NM}(s^2)
\over s^2(s-\omega^2)}.
\end{eqnarray}
By repeating the same step as above, we get
\begin{eqnarray}
\Pi^{\rm NM}(\omega^2)=c_{-1}+c_0\omega^2+{\omega^4\over\pi}
\int_{0+}^\infty \,ds\,{{\rm Im}\Pi^{\rm NM}(s)
\over (s-i\varepsilon)^2
(s-\omega^2-i\varepsilon)}.
\label{eqc.5}
\end{eqnarray}
Since ${\rm Im}\Pi^{\rm NM}(s)={\rm Im}s\Pi_1(s)$, (\ref{eqc.5})
is nothing but the relation obtained by multiplying $\omega^2$ to
both sides of (\ref{eqc.3}).  Therefore it is trivially correct that,
regardless of which sum rules we start with, $\Pi_1^{\rm NM}$ 
or $\Pi^{\rm NM}$, we get the same FESR.
This also applies to  $\Pi_1$ and $\Pi$ in the vacuum,
in which case $c_{-1}=0$ and $c_0=\Pi_1(0)$.

Next we explicitly demonstrate how (\ref{eq3.8}) and the FESR 
for $\Pi_1$ and $\Pi$ are obtained
in the vacuum, making clear the implicit assumptions
behind.
As is shown in (\ref{eqc.4}),
the QSR for $\Pi_1(q^2)$ is
\begin{eqnarray}
\Pi_1^{\rm OPE}(q^2)=\Pi_1(0)+{q^2 \over \pi}\int_{0+}^\infty\,ds\,
{ {\rm Im}\Pi_1(s) \over s(s-q^2)}.
\label{eqc.6}
\end{eqnarray}
Using the explicit forms in (\ref{eq3.6}) and (\ref{eq3.7}), we get
\begin{eqnarray}
-{1 \over 8\pi^2}{\rm ln}{-q^2\over \Lambda^2} +{A\over q^4}
+{B\over q^6}={F'\over m_V^2-q^2}-{F'\over m_V^2}
-{1\over 8\pi^2}{\rm ln}
{s_0-q^2\over -q^2}+\Pi_1(0)
\label{eqc.7}
\end{eqnarray}
where we explicitly introduced a finite scale $\Lambda^2$ 
which specifies a renormalization scheme in the OPE side, and
the integral in the right hand side converges due 
to the subtraction
introduced in the dispersion relation.  We rewrite (\ref{eqc.8}) as
\begin{eqnarray}
{1 \over 8\pi^2}{\rm ln}{s_0-q^2\over -q^2} +{A\over q^4}
+{B\over q^6}={F'\over m_V^2-q^2}+\left\{ -{F'\over m_V^2}
-{1\over 8\pi^2}{\rm ln}
{\Lambda^2\over s_0}+\Pi_1(0)\right\}.
\label{eqc.8}
\end{eqnarray}
By comparing the large $-q^2$-behavior in both sides
of (\ref{eqc.8}), one sees that $\{\cdots\}$ piece (constant terms)
in the right hand side should vanish, which is the 
consistency requirement
for this sum rule.  This is the implicit assumption which leads to
the sum rule in (\ref{eq3.8}).
In the Borel sum rule analysis, this mentioning is 
irrelevant, since $\{\cdots\}$ piece simply disappears 
after the Borel
transform.  Under this condition we get three FESRs' for the three
unknowns $F$, $m_V$ and $s_0$, by expanding both sides of 
(\ref{eqc.8})
with respect to ${1\over q^2}$ and comparing the coefficients of 
$1/q^2$, $1/q^4$ and $1/q^6$.

If one start with $\Pi(q^2)$ 
using the dispersion relation (\ref{eqc.5})
what we get is simply the relation
which is obtained by multiplying $q^2$ to both sides of (\ref{eqc.8}).
We then get the same FESRs'.
 
As is mentioned in the text, whether one applies Borel transform
to $\Pi_1$ or $\Pi$ causes numerical difference, especially because
the polynomial terms disappear by the Borel transform.  But
the FESR gives the same result regardless of which one we start with.

\newpage

\centerline{\bf Figure captions}

\vspace{0.5cm}

\begin{enumerate}

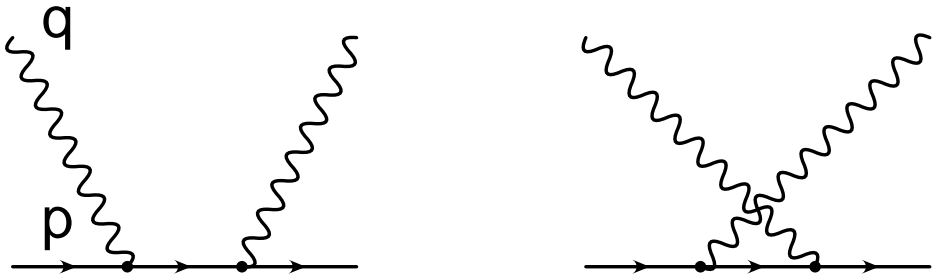
\item[{\bf Fig. 1}]
Born diagram contribution to $T_{\mu\nu}$.

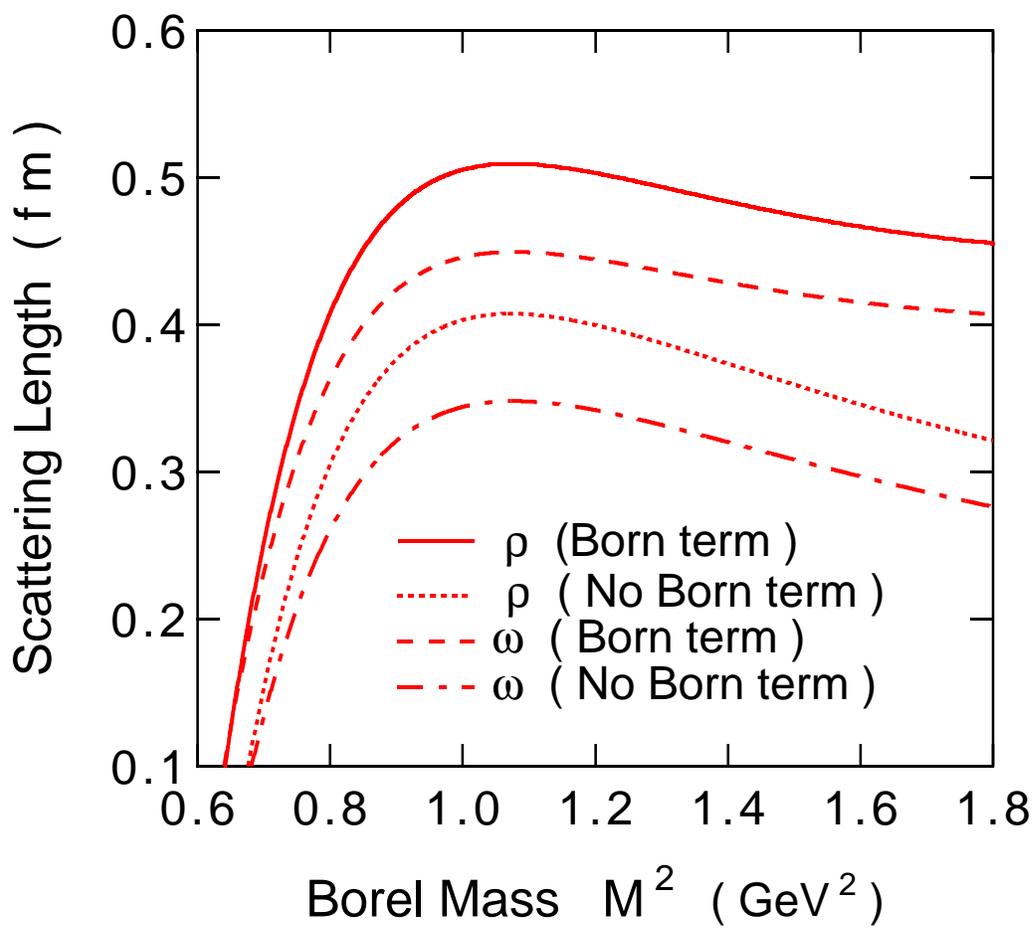
\item[{\bf Fig. 2}]
Borel surves for the $V-N$ ($V=\rho, \omega$) 
spin-isospin averaged
scattering lengths in the DBSR.  The results with and without 
explicit Born term (cases (i) and (ii)) are shown.

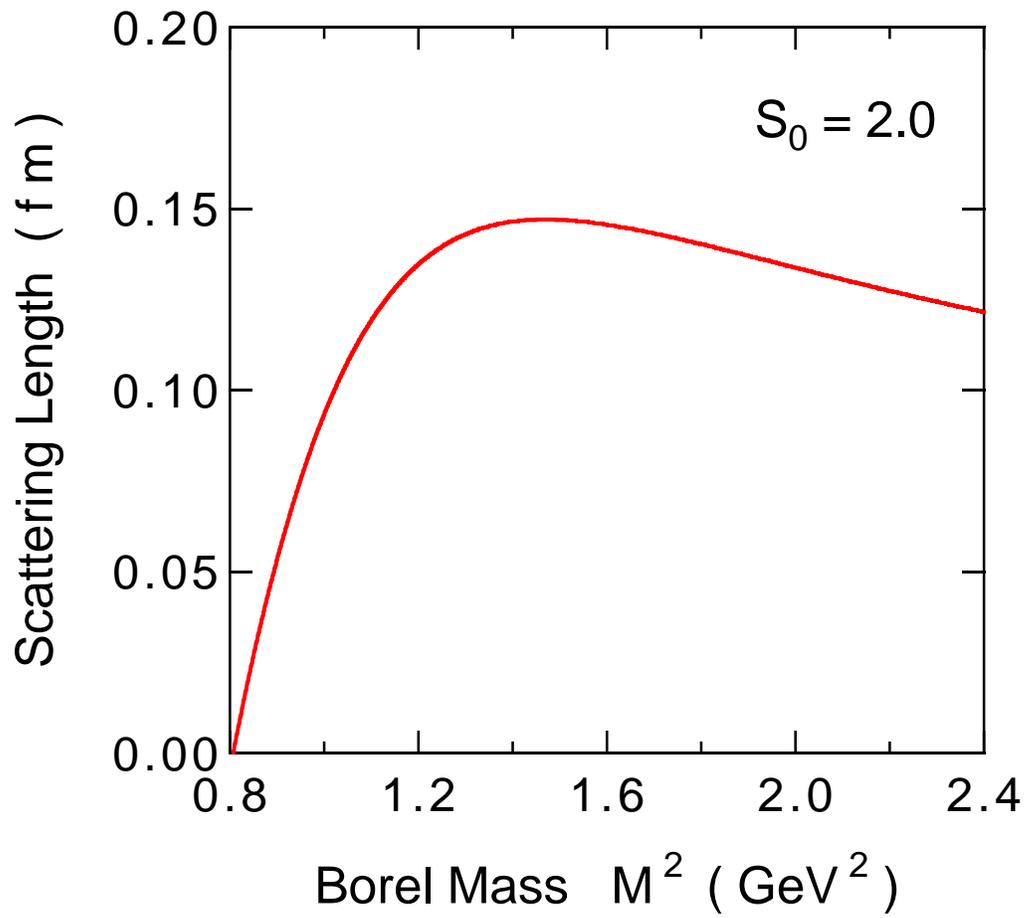
\item[{\bf Fig. 3}]
Borel curve for the $\phi-N$ spin-averaged scattering length
in the DBSR.

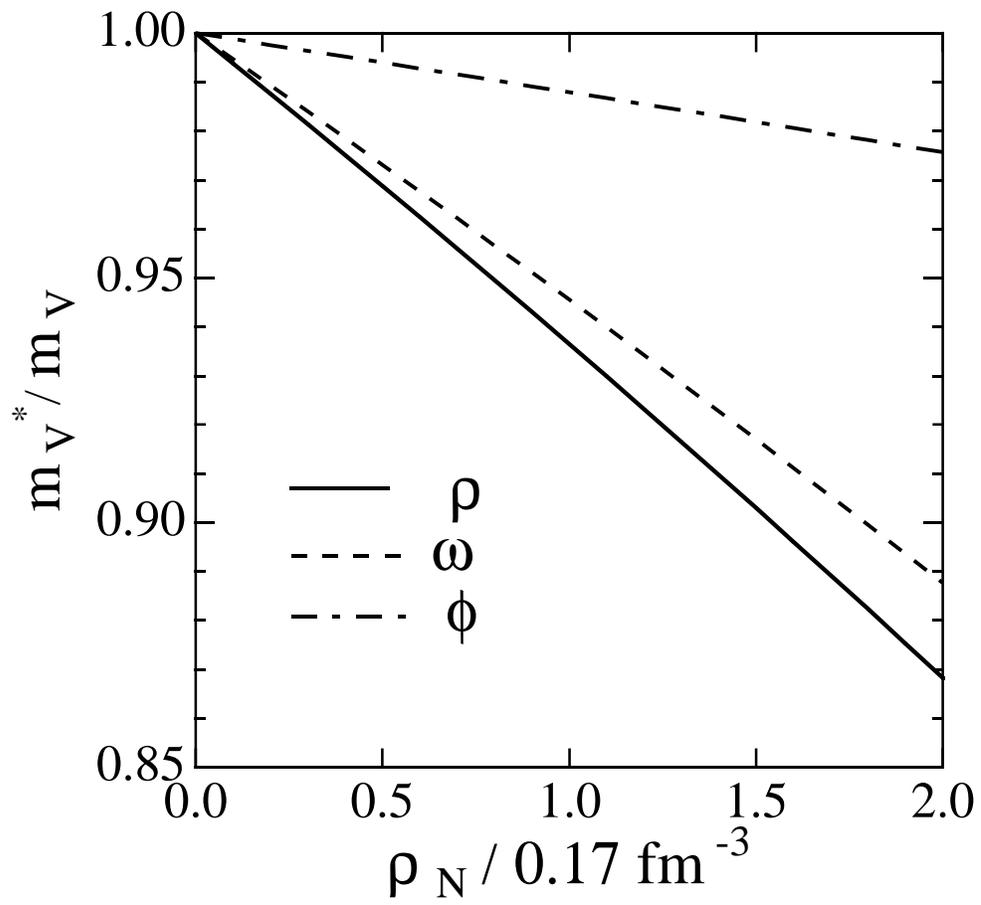
\item[{\bf Fig. 4}]
Vector meson masses in the nuclear medium ($m_V^*$)  
normalized by their vacuum values ($m_V$)
as a function of the nucleon density $\rho_N$ obtained from the 
scattering lengths $a_V$
in the linear density approximation.

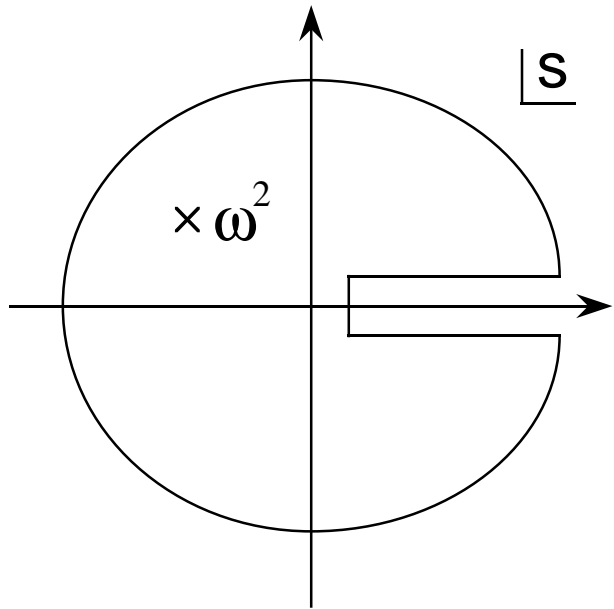
\item[{\bf Fig. 5}]
Closed path $C_1$ for the dispersion integral in (\ref{eqc.3}).

\end{enumerate}


\newpage
\begin{thebibliography}{99}
\bibliographystyle{unsrt}

                                                              
\setlength{\itemsep}{0.0in}


\bibitem{HL} T. Hatsuda and Su-H. Lee,  
Phys. Rev. {\bf C46}, R34 (1992).

\bibitem{Koike} Y. Koike,  
Phys. Rev. {\bf C51}, 1488 (1995).

\bibitem{KM} This observation was first applied to
the NN forward scattering amplitude by 
Y. Kondo and O. Morimatsu,
Phys. Rev. Lett. {\bf 71}, 2855 (1993).

\bibitem{HLS} T. Hatsuda, Su-H. Lee 
and H. Shiomi,  Phys. Rev. {\bf C52}, 3364 (1995).

\bibitem{SVZ} M.A. Shifman, 
A.I. Vainstein and V.I. Zakharov, 
Nucl. Phys. {\bf B147}, 385 (1979);
L.J. Reinders, H. Rubinstein and S. Yazaki, Phys. Rep. {\bf
127}, 1 (1985).

\bibitem{DL} 
E.G. Drukarev and E.M. Levin, Nucl. Phys. {\bf A511}, 679 (1990).


\bibitem{BS} A.I. Bochkarev 
and M.E. Shaposhnikov, Nucl. Phys. {\bf B268},
220 (1986).

\bibitem{err} 
In ref.\cite{Koike}, there was an error in 
the definition of 
the spin-averaged combination and the spin-averaged
scattering lengths.  
Since $\mbox{\boldmath $p$}=\mbox{\boldmath $q$}
=0$ was taken in \cite{Koike},
$T_2$ did not contribute and thus there was no error in $T$.
But the spin-averaged scattering lengths was defined 
as 1/3 of the correct
(present) one.


\bibitem{HKL} T. Hatsuda, Y. Koike and Su H. Lee, 
Nucl. Phys. {\bf B394},
221 (1993).

\bibitem{CHKL} S. Choi, T. Hatsuda, Y. Koike and S.H. Lee,
Phys. Lett. {\bf B312}, 351 (1993).

\bibitem{Bj} 
See for example,
V. De Alfaro, S. Fubini, G. Furlan and 
C. Rossetti, {\it Currents in Hadron Physics} (North-Holland, 1973).

\bibitem{KMN} Y. Kondo, O. Morimatsu and Y. Nishino, 
Phys. Rev. {\bf C53},
1927 (1996).

\bibitem{JL} X. Jin and D.B. Leinweber, 
Phys. Rev. {\bf C52}, 3344 (1995).

\bibitem{FH} R.J. Furnstahl and T. Hatsuda, Phys. Rev. Lett.
{\bf 72}, 3128 (1994).

\bibitem{KM2} Y. Kondo and O. Morimatsu, Phys. Rev. Lett.
{\bf 72}, 3129 (1994).

\bibitem{Koi} Y. Koike, Phys. Rev. {\bf D48}, 2313 (1993).

\bibitem{EI} V.L. Eletsky and B.L. Ioffe, 
Phys. Rev. {\bf D47}, 3083 (1993).

\bibitem{Hatsuda} T. Hatsuda, ``Talk at the 25th INS 
International Symposium on Nuclear and Particle Physics with 
High-Intensity Proton Accelerators'', (Dec. 3-6, 1996, Tokyo, Japan),
nucl-th/9702002.

\bibitem{SMS} K. Saito, T. Maruyama and K. Soutome, 
Phys. Rev. {\bf C40}, 407 (1989); H. Kurasawa and T. Suzuki,
Prog. Theor. Phys. {\bf 84}, 1030 (1990); H.-C. Jean, J. Piekarewicz and
A.G. Williams, Phys. rev. {\bf C49}, 1989 (1994); 
H. Shiomi and T. Hatsuda, Phys. Lett. {\bf B334}, 281 (1994).


\bibitem{Chin}
S.A. Chin, Ann. Phys. (N. Y. ), {\bf 108}, 308 (1977).





\end{thebibliography}
\end{document}